\documentclass[12pt,twoside]{article}   
\usepackage[super,sort,comma]{natbib}
\usepackage{amsmath,amssymb,amsfonts}
\usepackage{algorithmic}
\usepackage{graphicx}
\usepackage{textcomp}
\usepackage{hyperref}

\usepackage{ragged2e}   
\usepackage{adjustbox}
\usepackage{booktabs}
\usepackage{amsmath}
\usepackage{booktabs,tabularx}
\usepackage{multirow}
\usepackage{xcolor}
\usepackage{siunitx}
\usepackage[ruled,vlined]{algorithm2e}
\usepackage{bm}

\usepackage[font=normalsize]{caption}
\usepackage{fancyhdr}		

\usepackage[section]{placeins}   %

\usepackage{graphicx}

\makeatletter \renewcommand\@biblabel[1]{$^{#1}$} \makeatother
\newcommand{\cen}[1]{\begin{center} #1 \end{center}}

\usepackage[mathlines]{lineno}

\usepackage{hyperref}
\hypersetup{ colorlinks,
    citecolor=blue,
    filecolor=blue,
    linkcolor=blue,
    urlcolor=blue
}

\usepackage{xcolor}

\definecolor{gray}{rgb}{0.6,0.6,0.6}
\definecolor{red}{rgb}{0.85,0,0}
\definecolor{green}{rgb}{0,0.85,0}
\definecolor{blue}{rgb}{0,0,0.85}
\definecolor{beige}{rgb}{0.92,0.87,0.78}
\usepackage[all]{hypcap}    

\begin{document}

\cen{\sf {\Large {\bfseries Foveated-Imaging Geometry CT Architecture and Seeded Diffusion Model Enabling Global Super-Resolution Reconstruction } \\ 
		
\vspace*{10mm} 
Wenxin Mo$^{1,2}$, Yingxian Xia$^{1,2}$, Yongle Yan$^{1,2}$, Hao Zhou$^{1,2}$, Li Zhang$^{1,2}$ and Hewei Gao$^{*1,2}$
}
$^{1}$Tsinghua University, Beijing 100084, China \\
$^{2}$Key Laboratory of Particle \& Radiation Imaging (Tsinghua University), Ministry of Education,China
\vspace{5mm}\\
Version typeset \today\\
}

\pagenumbering{roman}
\setcounter{page}{1}
\pagestyle{plain}
Correspondence: Hewei Gao, Department of EngineeringPhysics, Tsinghua University, Beijing 100084,China. email: hwgao@tsinghua.edu.cn \\

\begin{abstract}
\noindent {\bf Background:} For X-ray computed tomography (CT), a smaller pixel size of detector, in general, leads to higher spatial resolution of the scanner, but inevitably results in increased system cost and data overhead in acquisition and processing. \\ 
{\bf Purpose:}  In order to achieve high-resolution (HR) CT imaging more resource-efficiently, we propose the Foveated-Imaging Geometry CT (FIGCT) architecture that integrates a local portion of HR data into low-resolution dominated acquisition, and develop a Diffusion Probabilistic FIGCT Super-resolution Reconstruction (DPFSR) framework that is able to generate global HR CT images over full field-of-view (FOV).\\
{\bf Methods:} The concept of FIGCT is first established, and its typical configurations are characterized according to the arrangement of HR data, with HR data fraction (HDF) and LR-to-HR detector pixel size ratio (LHR) introduced as two key indices. The proposed DPFSR incorporates local HR data information into intermediate clean image estimates in both the projection and image domains, during the intermediate steps of the reverse diffusion process. Such an additional step not only guides HR image generation from LR data but also improves data consistency between the clean-image estimates and the originally measured data.\\
{\bf Results:} Numerical simulations of FIGCT scans are conducted to analyze the spatially varying resolution under different HDFs and LHRs, followed by performance evaluation of DPFSR using the clinical dataset from an AAPM grand challenge and a set of swine lung CT data. The AAPM dataset included 5936 images, which were split into training, validation, and test sets at a ratio of 6:2:2. Preliminary results of Numerical simulations on FIGCT show that FIGCT provides high-precision CT images in the region of interest (ROI) corresponding to the HRdata,whilethespatialresolution deteriorates rapidly outside the ROI. With DPFSR, global HR reconstruction is achieved on AAPM grand challenge dataset and swine lung CT data, outperforming existing SR methods in terms of Learned Perceptual Image Patch Similarity, PSNR, and SSIM. \\
{\bf Conclusions:} This study demonstrates the feasibility of FIGCT as a resource-efficient CT imaging architecture for mixed-resolution data acquisition. By combining FIGCT with the proposed DPFSR algorithm, global HR CT reconstruction can be achieved while reducing the dependence on uniformly HR detector elements. The results suggest that FIGCT, together with HR-seeded diffusion reconstruction, provides a promising strategy for cost-effective and data-efficient HR CT imaging. \\

\end{abstract}
%
%
%
%
%
%
\section{Introduction}

High spatial resolution has long been a pursuit in CT imaging, as it directly contributes to improved diagnostic accuracy and the detection of subtle pathological changes \cite{Pontone14Radio}.
In practical implementation, high-resolution (HR) CT imaging typically relies on detectors with small pixel sizes and/or X-ray source with small focal spot, which could significantly increase system cost and data overhead in acquisition and processing\cite{yang2024empirical}. Obtaining HR images in a more resource-efficient manner is highly desirable, and numerous efforts have been made toward this goal. In the literature, existing approaches can be broadly categorized into hardware-based and algorithm-based methods.

From the hardware perspective, spatial resolution in CT is primarily determined by detector element size, X-ray focal spot size, and imaging geometry \cite{mp_resolution_impact}. In practice, the focal spot size is constrained by tube power and heat dissipation, making other components the primary means for improving resolution in conventional clinical CT. Several approaches have been proposed to enhance spatial resolution through system design without requiring uniformly high-resolution detectors. For example, Pan et al. proposed a geometry that increases projection sampling density by shifting the center of rotation \cite{xiaochuan_pan_spatial-resolution_2005}. Zhao et al. achieved super-resolution by combining subpixel-shifted acquisitions with iterative reconstruction \cite{Zhao_2023}. Haneda et al. proposed two off-center partial scans to exploit higher geometric magnification in peripheral regions \cite{haneda2022high}.

In many clinical scenarios, only specific regions are of interest, such as cardiac imaging and image-guided radiotherapy. In industrial CT, the required spatial resolution can also be direction-dependent, such as the nondestructive inspection of battery electrode sheets. Early multi-detector-row CT (MDCT) systems used detector elements of different sizes arranged along the $z$ direction, forming an adaptive array detector design. This configuration enabled high spatial resolution in the central slices while simultaneously providing a larger longitudinal field of view \cite{prokop_general_2003}. These observations suggest that non-uniform sampling geometries may better match task-specific imaging requirements while improving resource efficiency.

From the algorithmic perspective, most approaches aim to recover HR images from low-resolution (LR) inputs without requiring additional hardware, which is commonly referred to as super-resolution (SR). Classical SR techniques mainly include model-based iterative methods, such as prior-guided deconvolution\cite{stevenhigh,tilley2015model}, and learning-based methods, such as dictionary learning and sparse representation \cite{zhong2021preliminary,LI2016196}.

In the past years, an increasing number of deep learning methods have been developed for CT super-resolution (SR), where neural networks are trained to learn mappings from low-resolution (LR) to high-resolution (HR) images \cite{zhang2021self,chi2022ct,zhu2024super}. For example, Yu et al. employed SRCNN to enhance spatial resolution and preserve fine structures \cite{yu_computed_2017}, while Park et al. proposed a modified U-Net to map thick-slice images to thin-slice images \cite{Park_SR}. However, these methods may generate oversmoothed outputs.

More recently, denoising diffusion probabilistic models (DDPMs) have gained increasing attention \cite{DDPM,SR3,SRDiff,dolce}. DDPMs learn to progressively remove noise at different levels, enabling the transformation from Gaussian noise to an arbitrary data distribution. In super-resolution tasks, DDPMs have emerged as a powerful alternative to convolutional neural networks (CNNs) and generative adversarial networks (GANs), producing more realistic textures and fine details through their probabilistic sampling process. However, deep learning-based methods’ ability to recover true details beyond those contained in LR images remains questionable\cite{zhang21,tivnan_hallucination_2024,CircleGAN}.

To address these limitations, we propose a Foveated-Imaging Geometry CT (FIGCT) architecture, which integrates a local portion of HR data into an otherwise low-resolution-dominated acquisition. This design aims to achieve full-field-of-view (FOV) HR CT imaging while avoiding the cost and data burden associated with globally HR detectors. Based on this architecture, we further develop a Diffusion Probabilistic FIGCT Super-resolution Reconstruction (DPFSR) framework to generate global HR CT images from mixed-resolution measurements.

Specifically, we first establish the concept of FIGCT and characterize its typical configurations according to the arrangement of HR data. Two key indices, namely the HR data fraction (HDF) and the LR-to-HR detector pixel-size ratio (LHR), are introduced to describe the hybrid-resolution sampling geometry. We then formulate DPFSR as a seeded diffusion model-based SR reconstruction framework. During the intermediate steps of the reverse diffusion process, local HR information is incorporated into the intermediate clean-image estimates in both the projection and image domains. This seeded guidance strategy not only guides HR image generation from LR-dominated data but also improves data consistency between the clean-image estimates and the originally measured HR data. The proposed method is validated on the AAPM Low-Dose CT Grand Challenge dataset and a swine lung CT dataset. The \textbf{main contributions} are summarized as follows.

\begin{itemize}
	
	\item \textit{First}, we \textbf{present a new concept of CT imaging architecture}, i.e., FIGCT, which enables resource-efficient acquisition of HR data in task-specific regions or directions.

	\item\textit{Second}, we develop a seeded global super-resolution reconstruction framework enabling global super resolution reconstruction for FIGCT, i.e., DPFSR, which is the \textbf{first seeded diffusion model for SR} that incorporates local HR data information in image and projection domains.

	\item\textit{Third}, this is the \textbf{first attempt to demonstrate the effectiveness of FIGCT-type architecture for CT SR}, showing that its intrinsic HR components can support global SR reconstruction beyond conventional CT acquisitions.
\end{itemize}

\section{Methods}

\subsection{FIGCT-Type Architecture}
\subsubsection{The Concept of FIGCT}
In this work, FIGCT refers to a type of CT architecture that employs both HR and LR detector elements to acquire mixed-resolution data within a single CT scan. This can be implemented either by using the same detector type, such as a flat-panel detector or a photon-counting detector, operated with different binning modes, or by combining different detector types, such as energy-integrating detectors and photon-counting detectors. 

\begin{figure*}[ht]
	\begin{center}
			\includegraphics[width=\linewidth]{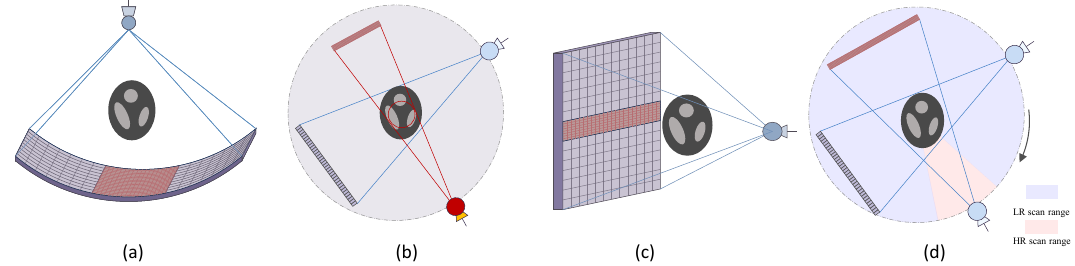}
	\end{center}
	\caption{FIGCT configuration categorization: 
		(a) FIGCT with different detector pixel sizes along the $x$ direction; 
		(b) D-FIGCT with a dual-source dual-detector configuration using different detector pixel sizes; 
		(c) Z-FIGCT with different detector pixel sizes along the $z$ direction; 
		(d) A-FIGCT using the same detector operated with different binning modes across projection views.}
	\label{fig:FIGCT_example} 
\end{figure*}

FIGCT can be classified into four categories according to the arrangement of HR detector elements as shown in Fig.~\ref{fig:FIGCT_example}. The first and most representative configuration, places HR detector elements at the center and LR ones at the periphery along the $x$ direction in a channel-wise manner. A similar mixed-resolution sampling pattern can also be realized by using a dual-source dual-detector configuration, referred as D-FIGCT. We also identify two generalized FIGCT configurations that may be useful in practice: mixed-resolution sampling is distributed along the $z$ direction (Z-FIGCT), with some rows of detectors having a smaller pixel size, and along the angular direction (A-FIGCT), with detector operated in different binning modes across projection views. 

Different FIGCT configurations are suited to different application scenarios. In the literature, D-FIGCT has been proposed for interior tomography\cite{luo_interior_2018} and Z-FIGCT has been applied to multi-row detector CT\cite{prokop_general_2003}. A-FIGCT has potential applications in nondestructive inspection of battery electrode sheets. 

In this study, we focus on the first FIGCT configuration in Fig.~\ref{fig:FIGCT_example}(a). This configuration naturally leads to a separation of the reconstruction into an interior HR region and an exterior LR region. In medical imaging, diagnostically relevant information is often concentrated in specific regions. Therefore, HR reconstruction can be selectively prioritized within these regions, while LR reconstruction in the exterior region provides anatomical context. Unless otherwise specified, FIGCT in this paper refers to this architecture.

To characterize the detector configuration, we define HR data fraction (HDF), denoted by $r$, and LR-to-HR detector pixel size ratio (LHR), denoted by $k$:
\begin{equation}
	r = \frac{S_h}{S_t},
\end{equation}
\begin{equation}
	k = \frac{\Delta d_l}{\Delta d_h},
\end{equation}
where $S_h$ denotes the area of the HR detector region and $S_t$ denotes the total detector area. For A-FIGCT, $S_h$ and $S_t$ correspond to the angular range covered by HR sampling and the full angular range, respectively. $\Delta d_l$ and $\Delta d_h$ denote the pixel sizes of the LR and HR detector elements, respectively.
\subsubsection{Initial Reconstruction Algorithm}
Due to the mixed detector resolutions in FIGCT, direct application of conventional filtered backprojection (FBP) is inappropriate, since the HR and LR measurements are sampled on different detector. To handle this resolution mismatch, Multiplexing Split FBP (MS-FBP) is adopted to obtain an initial low-resolution reconstruction from FIGCT data~\cite{Xia20}. The detailed procedure of MS-FBP is summarized in Algorithm~\ref{alg:mfbp}.

For a sinogram $P_{\mathrm{FI}}$ acquired from FIGCT, we first decompose it according to the detector configuration as
\begin{equation}
	P_{\mathrm{FI}} = \mathcal{C}(P_{L_1}, P_H, P_{L_2}),
\end{equation}
where $P_H$ denotes the projection data acquired by the HR detector elements, and $P_{L_1}$ and $P_{L_2}$ denote the projection data acquired by the LR detector elements on the two sides. The operator $\mathcal{C}(\cdot)$ represents concatenation along the detector coordinate.
\begin{algorithm}[t]
	
	\caption{MS-FBP for FIGCT Reconstruction}
	\label{alg:mfbp}
	\KwIn{$P_{L_1}, P_{L_2}$: LR projections, $P_H$: HR projection}
	
	$P_{H_{\text{int}}} \leftarrow \mathcal{C}\bigl(\mathcal{I}(P_{L_1}),\, P_H,\, \mathcal{I}(P_{L_2})\bigr)$ \tcp*[r]{concatenation}
	$P_{L_{\text{int}}} \leftarrow \mathcal{C}\bigl(P_{L_1},\, \mathcal{D}(P_H),\, P_{L_2}\bigr)$\;
	$\tilde{P}_{H_{int}} \leftarrow w \odot P_{H_{\text{int}}}*h_H$ \tcp*[r]{weighting and filtering}
	$\tilde{P}_{L_{int}} \leftarrow w \odot P_{L_{\text{int}}}*h_L$\;
	$\tilde{P'}_{H} \leftarrow \textbf{Int}(\tilde{P}_{H_{int}})$\;
	$\tilde{P'}_{L_1},\ \tilde{P'}_{L_2} \leftarrow \textbf{Ext}(\tilde{P}_{L_{int}})$\;
	$f \leftarrow \int_{0}^{2\pi} \mathcal{C}(\tilde{P'}_{L_1},\, \tilde{P'}_{H},\, \tilde{P'}_{L_2}) d\theta$ \tcp*[r]{back-projection step in FBP}
	\KwOut{$f$ (reconstructed image)}
	
	\noindent\rule{\linewidth}{0.4pt}
	\vspace{-4pt}
	
	\footnotesize
	\noindent\parbox{\linewidth}{
		\textit{Note:} $\mathcal{I}(\cdot)$ denotes interpolation from the LR grid to the HR grid, $\mathcal{D}(\cdot)$ denotes downsampling from the HR grid to the LR grid, $*$ denotes convolution, $w$ denotes the geometric weighting term used in FBP, and $h_H$ and $h_L$ denote the FBP filtering kernels for the HR and LR data, respectively.
	}
\end{algorithm}
Accordingly, we define the interior and exterior extraction operators as
\begin{equation}
	P_H = \mathbf{Int}(P_{\mathrm{FIG}})
	\qquad 
	(P_{L_1}, P_{L_2}) = \mathbf{Ext}(P_{\mathrm{FIG}}),
\end{equation}
which extract the HR and LR projection components, respectively. 


\subsubsection{FIGCT Characteristics}

\begin{figure}[t]
	\begin{center}
			\includegraphics[width=\linewidth]{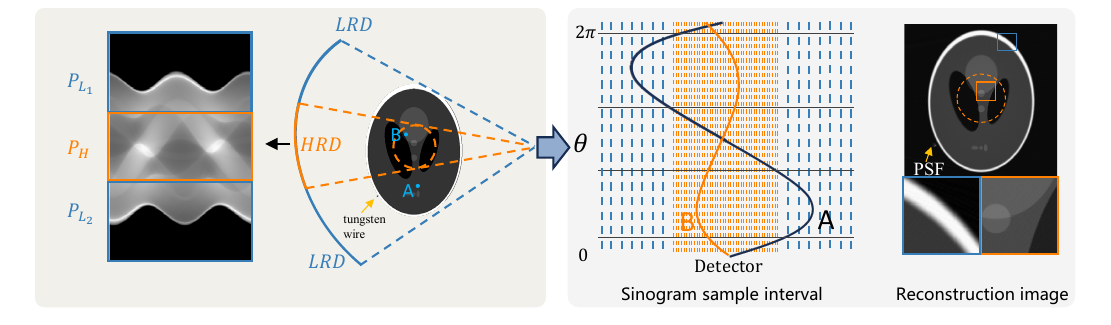}
	\end{center}
	\caption{FIGCT sampling characteristics. The HR detector region provides full-angular sampling for point B in the interior region, whereas point A in the exterior region is sampled by the HR detector region only over a limited angular range. This sampling behavior leads to spatially varying resolution in the reconstructed image.}
	\label{fig:FIGCT} 
\end{figure}

FIGCT projection data exhibit foveated characteristics, as shown in Fig.~\ref{fig:FIGCT}. 
For an FIGCT scan, mixed resolution sampling are observed: a point located in the exterior region (denoted by A), is only partially sampled over a limited angular range, whereas a point located in the interior region (denoted by B), is sampled by the HR detectors over the full angular range. This inherent sampling property enables HR reconstruction in the interior region, while producing direction-dependent radial blurring in the exterior region, where spatial resolution gradually deteriorates with increasing distance from the center. Consequently, the reconstructed image also exhibits a foveated appearance with spatially varying resolution. 

\begin{figure*}[ht]
	\begin{center}
			\includegraphics[width=\linewidth]{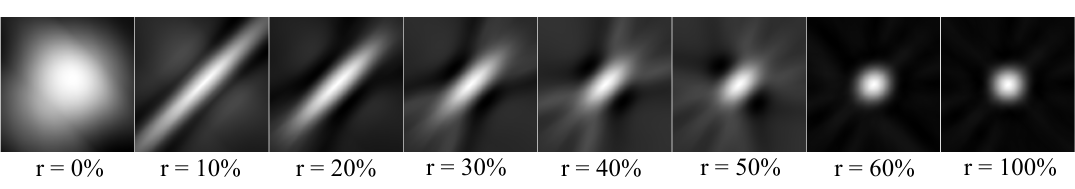}
	\end{center}
	\caption{PSFs of an exterior point in FIGCT with different HDFs, $r$, at a fixed LHR of $k=4$. As $r$ increases, the PSF gradually transitions from anisotropic to approximately isotropic.}
	\label{fig:PSF} 
\end{figure*}

Reconstruction of a tungsten wire with a diameter of 50~$\mu$m positioned at the bottom-left of the phantom in Fig.~\ref{fig:FIGCT} reveals the point spread function (PSF) of FIGCT scans with various HDFs, $r$. As shown in Fig.~\ref{fig:PSF}, the central HR detector improves spatial resolution even in exterior regions; however, this improvement is anisotropic and limited to specific directions. This behavior essentially corresponds to a limited-angle HR sampling problem \cite{Zhang_2025}. When the tungsten wire is fully covered by the HR detector region during rotation, the resulting PSF becomes approximately isotropic and approaches that of a system with global HR detectors.

These observations suggest that the HR projection segment in FIGCT contains not only interior HR information but also direction-dependent HR structural information from exterior regions. Therefore, although MS-FBP in Algorithm~\ref{alg:mfbp} provides only an initial low-resolution reconstruction, the HR projection measurements retain additional high-frequency information that can be further exploited for subsequent global SR reconstruction.
\subsection{DPFSR Super-Resolution Reconstruction Framework}
\begin{figure*}[ht]
	\begin{center}
			\includegraphics[width=1\linewidth]{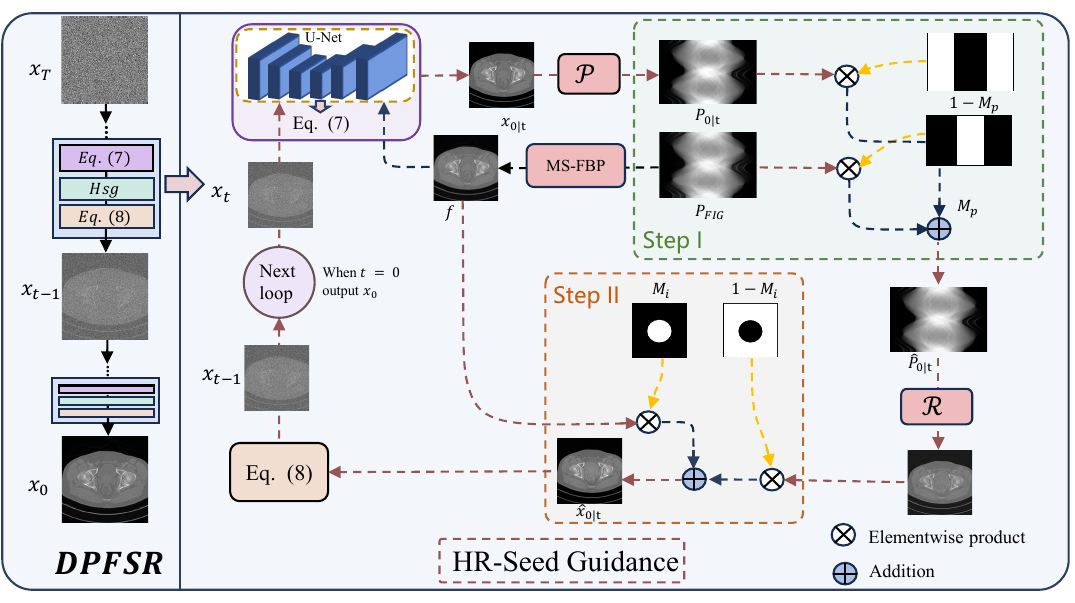}
	\end{center}
	\caption{Overall pipeline of DPFSR. An initial LR image is first reconstructed using MS-FBP. During each reverse diffusion step, HR-seed guidance is applied to the intermediate clean-image estimate in both the projection and image domains. The measured HR projection segment is injected using the projection-domain mask $M_p$, while the reliable interior region from the initial reconstruction is preserved using the image-domain mask $M_i$. $\mathcal{P}(\cdot)$ denotes the forward-projection operator with a full HR detector geometry, and $\mathcal{R}(\cdot)$ denotes the FBP reconstruction operator.}
	\label{fig:DPFSR} 
\end{figure*}

Since the exterior region is partially sampled by the HR detector, HR structural information from the exterior region is embedded in the HR projection segment. Our goal is to reconstruct global SR images from given mixed-resolution projections, $P_H, P_{L_1}$ and $ P_{L_2}$. Although $P_H$ has been used in MS-FBP to reconstruct $f$, its high-frequency information is implicitly mixed with LR data during interpolation, filtering, and backprojection, and therefore may not be fully preserved in the image-domain condition. Inspired by the recent successes of DDPMs \cite{Repainted,dolce}, we propose DPFSR based on conditional DDPMs, where \(P_H\) is introduced as an explicit projection-domain condition during the reverse diffusion process. The overall framework is shown in Fig.~\ref{fig:DPFSR}. It is worth noting that the proposed DPFSR can also be readily extended to A-FIGCT and D-FIGCT configurations.

\subsubsection{Conditional DDPM}
Conditional DDPM (cDDPM) is a diffusion model conditioned on a guidance image, consisting of a forward diffusion process and a reverse denoising process. The forward process gradually adds Gaussian noise to an image, generating a sequence
$x_0 \rightarrow x_1 \rightarrow \cdots \rightarrow x_T$, where noise increases from the clean image $x_0$ to pure Gaussian noise $x_T$:

\begin{equation}
	\begin{aligned}
		q(x_t|x_0) &= \mathcal{N}(x_t;\sqrt{\gamma_t}x_0,(1-\gamma_t)I), \\
		\gamma_t &= \prod_{s=1}^{t}\alpha_s, ; \alpha_t = 1-\beta_t
	\end{aligned}
\end{equation}

In the reverse process, cDDPM generates HR images by modeling
$p_\theta(x_{t-1}|x_t, f)$ conditioned on the noisy image $x_t$ and the corresponding LR image $f$.
With a neural network predictiang the noise $\tilde{\epsilon}_\theta(x_t, f, t)$, so $x_{t-1}$ can be estimated as:

\begin{equation}	
	x_{t-1} = \frac{1}{\sqrt{\alpha_t}} \left(x_t - \frac{1-\alpha_t}{\sqrt{1-\gamma_t}} \tilde{\epsilon}_\theta(x_t, f, t)\right) + \sigma_{t}z
\end{equation}

where $z \sim \mathcal{N}(0,I)$. Equivalently, the reverse sampling step can also be written as follow.

The intermediate clean-image estimate $x_{0|t}$ can be predicted from $x_t$:

\begin{equation}
	x_{0|t} = \frac{1}{\sqrt{\gamma_t}}\Big(x_t - \sqrt{1-\gamma_t} \tilde{\epsilon}_\theta(x_t, f, t)\Big)
	\label{eq:pred}
\end{equation}

Following \cite{SR3}, the posterior sample of $x_{t-1}$ given $x_0$ and $x_t$ is:

\begin{equation}
	x_{t-1} =
	\frac{\sqrt{\gamma_{t-1}} \beta_t}{1-\gamma_t}{x}_{0}
	+
	\frac{\sqrt{\alpha_t}(1-\gamma_{t-1})}{1-\gamma_t}x_t
	+
	\sigma_t z
\end{equation}
where $\sigma_t^2 = \frac{(1-\gamma_{t-1})(1-\alpha_t)}{1-\gamma_t}$.

\subsubsection{HR-Seed Guidance}
In DPFSR, we introduce an HR-seed guidance (\textbf{Hsg}) step, which injects the measured HR projection segment \(P_H\) into the intermediate clean-image estimate \(x_{0|t}\). This enables the reverse diffusion process to progressively refine the SR image with projection-domain HR information as an explicit seed.

In the inference stage of DPFSR, when $ t_s \geq t \geq t_c$, the intermediate clean-image estimate predicted by Eq.~(\ref{eq:pred}) is followed by the \textbf{Hsg} process, which consists of two steps:

Step I: clean image estimates $x_{0|t}$ is forward-projected with a global HR detector geometry and the The corresponding HR projection segment is then replaced with the measured HR projection segment \(P_H\); 

\begin{equation}
	{\hat{P}_{0|t}} = (1-M_p) \odot \mathcal{P}(x_{0|t}) +  M_p \odot P_{FI}
\end{equation}
where $ M_p$ represents the mask of the HR projection segment. $t_s$ is the start step and $ t_c $ is the end step of $\textbf{Hsg}$.

Step II: preserving the reliable interior region from the initial reconstruction \(f\), since the interior region in \(f\) is fully sampled by the HR detector segment and is therefore more reliable.
\begin{equation}
	{\hat{x}_{0|t}} = (1-M_i) \odot \mathcal{R}({\hat{P}_{0|t}}) +  M_i\odot f
	\label{eq:M_image}
\end{equation}
where $ M_i$ represents the mask of the interior region of the image domain. $\mathcal{P}(\cdot)$ denotes the forward projection operation with a full HR detector.  $\mathcal{R}(\cdot)$ denotes the FBP reconstruction operator. The overall steps of DPFSR are summarized in Algorithm~\ref{alg:refine}. The forward-projection and FBP operations are implemented using CTLIB\cite{xia2021magic}.

During the preparation of this work, the authors used ChatGPT for language polishing and grammar refinement. After using this tool, the authors reviewed and edited the content as needed and take full responsibility for the content of the publication. The swine lung CT study was approved by the Institutional Animal Care and Use Committee (IACUC) of Fuwai Hospital, Chinese Academy of Medical Sciences (approval No. 0108-1/3-1/1-HX(X)-61), and all procedures were performed in accordance with the approved institutional guidelines.

\begin{algorithm}[t]
	\caption{DPFSR}
	\label{alg:refine}
	\KwIn{$P_{L_1}, P_{L_2}$: LR projections, $P_H$: HR projection, $f$: reconstructed LR image,
		$\tilde{\epsilon}_\theta$: Adjusted denoiser network}
	Sample $x_T \sim \mathcal{N}(0, I)$ \tcp*[r]{diffusion sampling}
	\For{$t = T, T-1, \ldots, 1$}{
		$z \sim \mathcal{N}(0, I)$\;
		$x_{0|t} = \frac{1}{\sqrt{\gamma_t}}
		\left(
		x_t - \sqrt{1 - \gamma_t}\, \tilde{\epsilon}_\theta  (x_t, f, t)
		\right)$\;
		if $ t \leq t_s$ \& $ t \geq t_c $ \\
		\quad	$ \hat{x}_{0|t} = \textbf{Hsg}\bigl(x_{0|t}, P_H, f\bigr) $\;
		else\\
		\quad	$ \hat{x}_{0|t} = {x}_{0|t}$\;
		
		$x_{t-1} =
		\frac{\sqrt{\bar{\gamma}_{t-1}}\,\beta_t}{1 - \gamma_t}\,\hat{x}_{0|t}
		+ \frac{\sqrt{\alpha_t}\,(1 - \gamma_{t-1})}{1 - \gamma_t}\,x_t
		+ \sigma_t z$\;
		
	}
	\KwOut{$x_0$: SR image}
\end{algorithm}

\subsubsection{Network Optimization and Noise Schedule}

In this study, we employ the U-Net architecture proposed by Saharia et al.~\cite{SR3} as the backbone of the proposed framework. The initial LR image $f$, reconstructed by MS-FBP, is concatenated with the noisy input $x_t$ along the channel dimension to condition the model. Similar to cDDPM~\cite{DDPM}, we use the $\ell_1$ loss as the training objective:
\begin{equation}
	\mathcal{L} =
	\mathbb{E}_{x_0, f, \epsilon, t \in [1,T]}
	\left[
	\left\|
	\epsilon_\theta(x_t, f, t)-\epsilon
	\right\|_1
	\right].
\end{equation}

For the training schedule, we set $T=2000$. A linear noise schedule is adopted, where the noise variances $\beta_1,\beta_2,\cdots,\beta_t,\cdots,\beta_T$ increase uniformly over the diffusion steps.

\subsection{Image Quality Metrics}
We use Peak Signal-to-Noise Ratio (PSNR),Structural Similarity Index Measure (SSIM) and Learned Perceptual Image Patch Similarity (LPIPS)\cite{LPIPS} as quantitative metrics to compare our method with others.

PSNR quantifies the pixel-level fidelity of a reconstructed image to a reference image, but it favors smoother images, while SSIM tends to miss subtle details and fine structures. In order to align the image quality assessment with human visual perception, we further employ LPIPS\cite{LPIPS} to evaluate the perceptual differences between images, which measures image difference in deep feature space of pretrained neural networks and more consistent withe human visual perception. A lower LPIPS value indicates greater similarity to the HR image. AlexNet was selected as backbone network for computing LPIPS. 

For a fair comparison, before metric computation, the interior region of each SR image \(x_{\mathrm{sr}}\) is replaced with the corresponding region from the MS-FBP reconstruction \(f\).
\begin{equation}
	\hat{x}_{\mathrm{sr}} = f \odot M_i + x_{\mathrm{sr}}\odot (1 - M_i)
	\label{eq:mask}
\end{equation}

\section{Experiments and Results}
\subsection{FIGCT Simulation Setup}
We used the geometry shown in Fig.~\ref{fig:FIGCT} for FIGCT simulation. In the FIGCT geometry, the source-to-detector distance (SDD) and source-to-isocenter distance (SID) were set to 1000 mm and 625 mm, respectively. The element size of HR detector was  0.75 mm and LHR $k$ was fixed at 4 after the FIGCT reconstruction analysis described in Section \ref{sec:sys}. A total of 720 fan-beam projections were uniformly sampled over 360°. When $ r = 100\%$, the number of detector elements in the HR detectors was 1120. For other values of $r$, adjacent HR detector elements were proportionally merged to form LR detector elements, while maintaining a constant FOV. 

In simulation, monoenergetic photons with an energy of 70 keV were incident on each HR detector element, with a total photon count of $10^8$ per pixel. For the LR detectors, the photon count was set to four times that of the HR detectors to account for the larger detector element size. After attenuation through the phantom, the transmitted photons were recorded at the detector, and the raw signal was generated by applying Poisson noise, with additional blur applied in the LR detector.

The simulation phantom, with data size of $512 \times 512$, was derived from clinical images provided by the AAPM Low Dose CT Grand Challenge dataset \cite{LDCT}. As an illustrative example, when $r=20\%$, the simulation produces a HR projection segment with size of $112 \times 720$, along with two LR projection segments of size $ 126 \times 720$ each. To preserve the high spatial resolution in the interior region, the initial reconstructed image was maintained at a matrix size of $512 \times 512$.


\subsection{Network Training and dataset}
We trained and evaluated DFPSR on NVIDIA Quodra RTX 8000 GPU. The noise schedule of cDDPM was linearly increased from  1e-6 to 1e-2. During DPFSR training, the batch size was set to just 4; the learning rate was initially warmed up and subsequently fixed at 1e-4, and a dropout rate of 0.2 was consistently applied throughout. The early stopping strategy was employed to prevent overfitting.

Datasets were generated under FIGCT configurations with $r = 20\%$ only using full dose images in AAPM dataset for methods comparison, while another datasets was generated  under conventional CT configurations with full LR detectors for geometry comparison. To evaluate the SR performance on A-FIGCT architecture, a datasets under A-FIGCT configurations with $r = 10\%$ was also generated. 
\subsection{Reference Methods}
In this study, DPFSR was compared with several state-of-the-art SR methods, including U-Net~\cite{U-net}, EDSR~\cite{EDSR}, GAN-CIRCLE~\cite{CircleGAN}, ESRGAN~\cite{esrgan}, and cDDPM~\cite{DDPM}. All comparison methods were trained and tested on the FIGCT dataset with \(r = 20\%\).

\subsection{SR Performance Comparison between FIGCT and Conventional CT Geometry}
Reference methods and ours were trained and tested on different datasets from two geometries separately. The same comparison was also conducted on the A-FIGCT architecture and conventional CT. Since DPFSR is designed specifically for FIGCT, it is not applicable to the conventional setting.
\subsection{Analysis of FIGCT Performance with Different Parameters}
\label{sec:sys}
\subsubsection{LHR}
\begin{figure*}[!t]
	\begin{center}
			\includegraphics[width=1\linewidth]{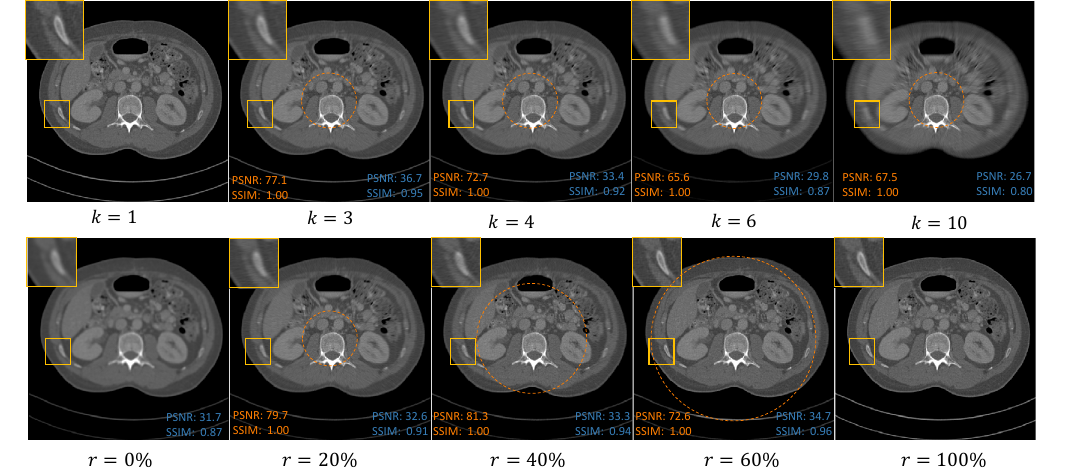}
	\end{center}
	\caption{Reconstruction result of different LHR $k$ with $r$ fixed to 20\% and different HDF $r$ with $k$ fixed to 4.. The display windows is [-500,1000] HU. The metric in the lower-left corner is computed between the circled region and the HR reference, while the one in the lower-right corner is calculated over the entire image.}
	\label{fig:rk-various}  
\end{figure*}

To investigate the effect of different LHR values $k$ on the reconstructed images, we fixed $r = 20\%$
and performed reconstructions with varying \(k\). The reconstruction results are shown in Fig.~\ref{fig:rk-various}. As $k$ increases, the exterior region becomes progressively blurred. When $k > 6$, low-frequency structures begin to lose clarity. As MS-FBP relies on interpolation, increasing $k$ leads to a degradation in the reconstruction quality of the HR interior region, where PSNR consistently decreases as $k$ increases. Therefore, in our geometry, we recommend $k \leq 6$ when preserving exterior-region image quality is important; in the subsequent simulations, we use $k = 4$.
\subsubsection{HDF}
We evaluate the reconstruction accuracy through numerical simulation. Results are presented in Fig.~\ref{fig:rk-various}. We observe that HR detectors could fully cover the phantom while $r = 60\%$. The orange circle denotes the interior region that is scanned by HR detectors for all view angles. The PSNR and SSIM performance within the circle and overall image are shown in the corner of each image. MS-FBP is theoretically incapable of achieving exact image reconstruction because it involves interpolation operations. Nevertheless, it enables high-quality reconstruction of the interior regions, yielding excellent quantitative metrics (PSNR $>$ 100 dB and SSIM close to 1). In the exterior region, the spatial resolution exhibits directional blurring outside the circle and gradually deteriorates with increasing distance from the center.

\subsection{SR Results}
\subsubsection{Comparison of DPFSR and Other Methods}
\begin{figure*}[!t]
	\begin{center}
			\includegraphics[width=1\linewidth]{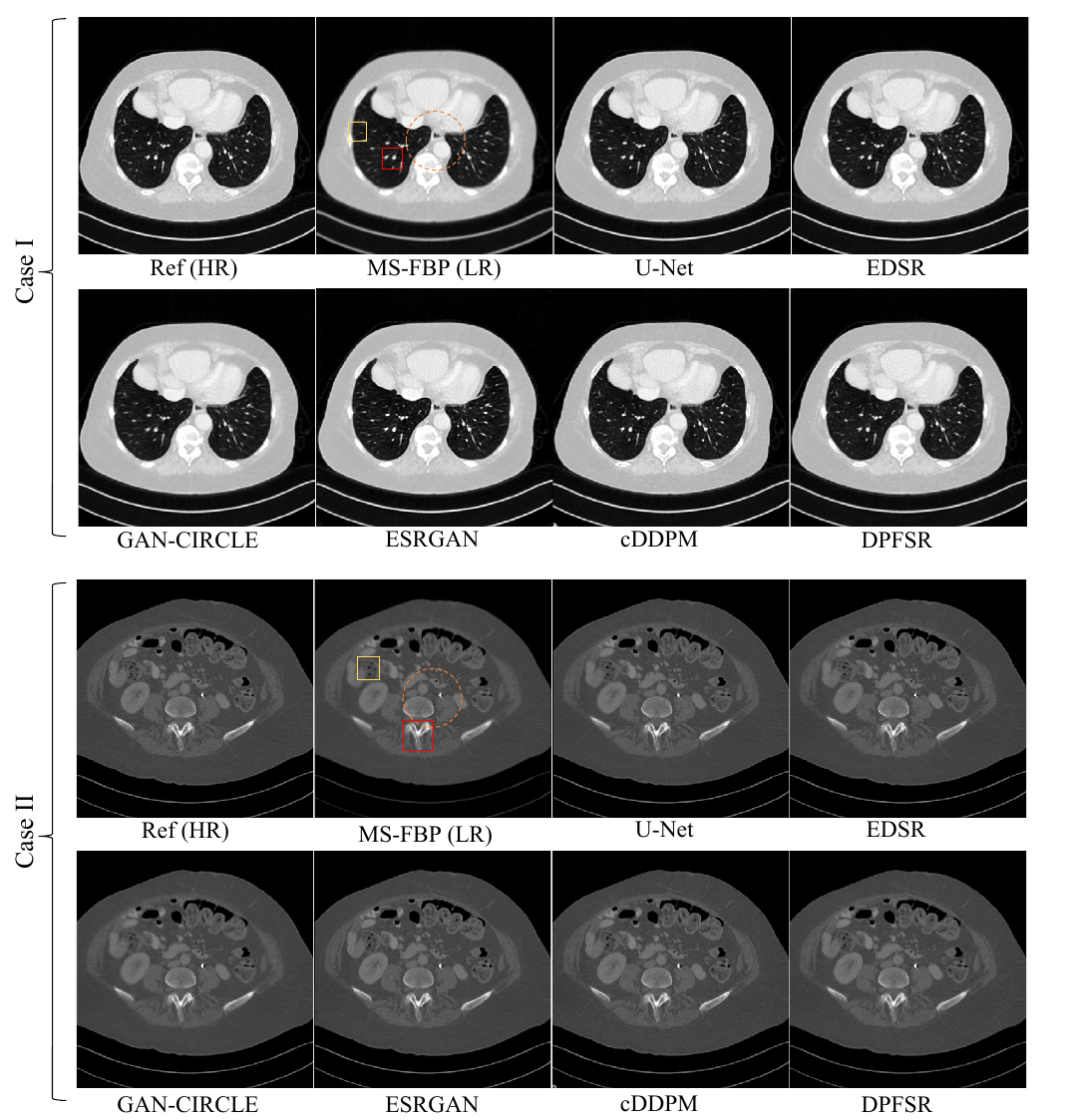}
	\end{center}
	\caption{Visual comparison of $r = 20\%$ case with different methods. The display windows is [-1000,250] HU for case I and [-500,1000] HU for case II. Region in orange circle represents HR area. ROIs in yellow and red rectangle is zoomed in at Fig.~\ref{fig:result_sim_roi}}
	\label{fig:result_sim} 
\end{figure*}

\begin{figure*}[!t]
	\begin{center}
			\includegraphics[width=1\linewidth]{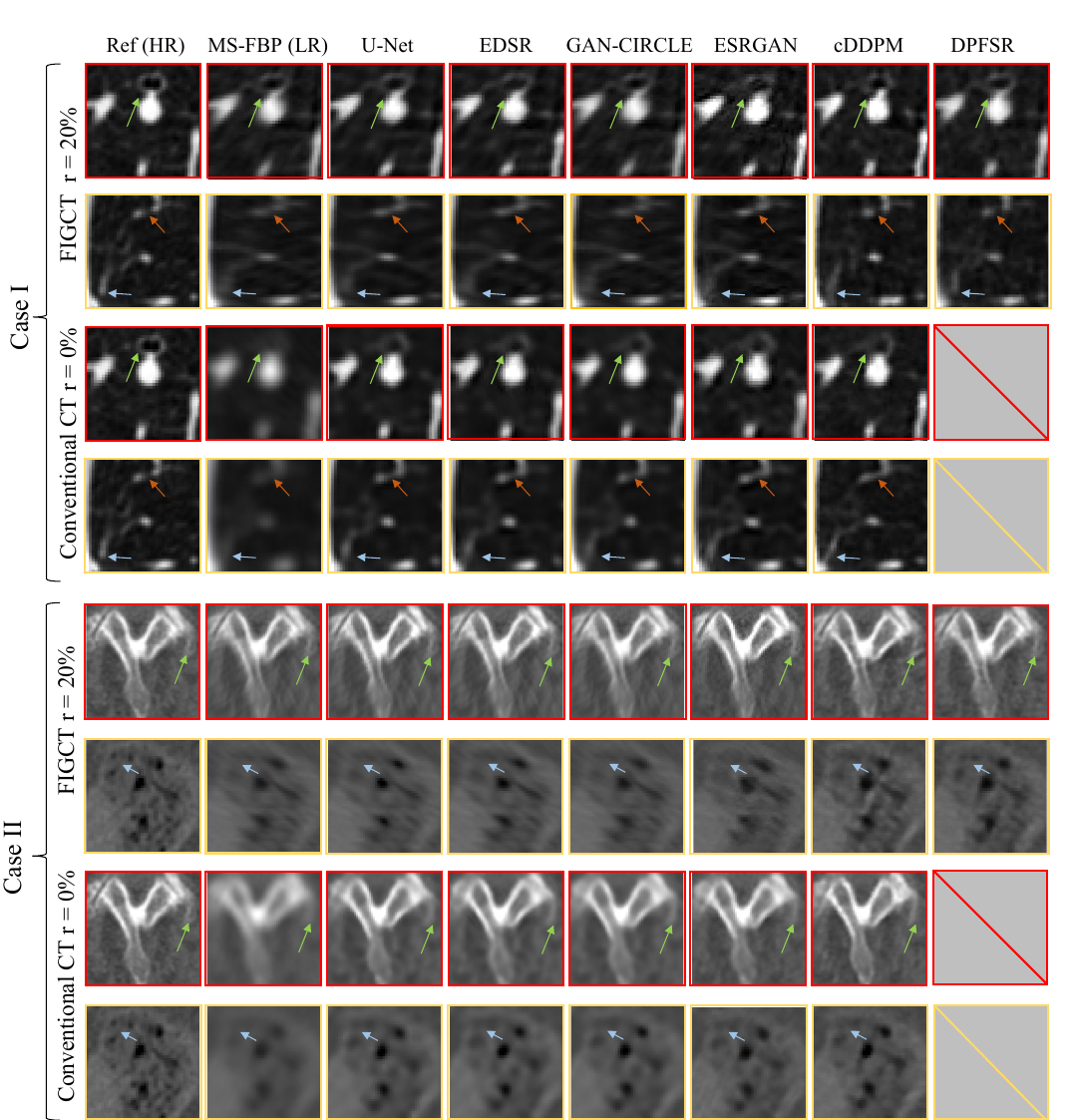}
	\end{center}
	\caption{Zoomed ROIs of cases in Fig.~\ref{fig:result_sim} of FIGCT $r = 20\%$ and conventional CT $ r = 0\%$ datasets across different methods. The display windows is [-1000,250] HU for case I and [-500,1000] HU for case II.}
	\label{fig:result_sim_roi} 
\end{figure*}
We compared DPFSR with reference methods on FIGCT $r = 20\%$ case.  Visual results are presented in Fig.~\ref{fig:result_sim} and Fig.~\ref{fig:result_sim_roi}. GAN-based and PSNR-oriented methods are capable of producing images with higher resolution, but the directional blur characteristics are still preserved in the SR images. Although DDPM can generate images with richer feature, its outputs still exhibit structural that do not exist on HR images and unable to restore some blurred structural details. In contrast, DPFSR achieves high-resolution global SR reconstruction, effectively recovering fine structures that are blurred in the LR images.

The quantitative results are presented in Table~\ref{tab:metrics}. DPFSR achieves the highest score in LPIPS with pleasing results in terms of PSNR and SSIM,  which is consistent with the visual comparisons. In other words, our method delivers consistent performance across the test set, producing images with anatomical structures that are better aligned with HR images.

\subsubsection{Comparison of FIGCT and Conventional CT Geometry}
To illustrate the role of the HR projection segment in FIGCT, we trained and evaluated different SR methods on datasets corresponding to the conventional CT case ($r=0\%$) and the FIGCT case ($r=20\%$). The quantitative results show that simply replacing conventional CT data with FIGCT data does not consistently improve the performance of existing SR methods. For the same method, models trained with the FIGCT $r=20\%$ case generally achieve lower PSNR than those trained with the conventional CT $r=0\%$ case. In terms of LPIPS, cDDPM and ESRGAN also exhibit degradation, whereas other methods show improvement.

This observation suggests that the HR information provided by FIGCT is not trivially exploitable by generic image-domain SR models. Since FIGCT reconstruction is shift-variant and exhibits spatially varying resolution, existing SR methods may not effectively handle the nonuniform degradation. Nevertheless, the visual results show that models trained with FIGCT $r = 20\%$ can recover richer details and sharper edges in certain regions, indicating that the HR projection segment indeed provides useful high-frequency information.

In contrast to existing methods, DPFSR explicitly incorporates the HR projection segment during the reconstruction process. As a result, the DPFSR model trained with FIGCT $r=20\%$ achieves higher scores across all metrics than the model trained with conventional CT $r=0\%$. These results indicate that the advantage of FIGCT does not come from the mixed-resolution data alone, but from effectively exploiting the retained HR projection measurements with a reconstruction method tailored to FIGCT.



\begin{table*}[t]
	\centering
	\caption{Quantitative comparison on the AAPM dataset and swine data under two settings $r = 20\%$ and $r = 0\%$ across different methods.}
	\renewcommand{\arraystretch}{1.6}
	\setlength{\tabcolsep}{2.5pt}
	\scriptsize   
	
	\begin{tabularx}{\textwidth}{@{\extracolsep{\fill}}lccc c ccc c ccc c ccc}
		\hline\hline
		& \multicolumn{7}{c}{AAPM}   & & \multicolumn{7}{c}{Swine data}     \\
		& \multicolumn{3}{c}{Conventional CT r = 0\%} & & \multicolumn{3}{c}{FIGCT r = 20\%}
		& & \multicolumn{3}{c}{Conventional CT r = 0\%} & & \multicolumn{3}{c}{FIGCT r = 20\%}                    \\
		\cline{2-4}\cline{6-8}\cline{10-12}\cline{14-16}
		& PSNR & SSIM  & LPIPS
		& & PSNR & SSIM  & LPIPS
		& & PSNR & SSIM  & LPIPS
		& & PSNR & SSIM  & LPIPS  \\
		DPFSR      
		&   -   &   -    &   -
		& & 39.9 & 0.959 & \underline{\textbf{0.043}}
		& &  -   &   -    &   -
		& & \underline{\textbf{34.2}} & 0.852 & \underline{\textbf{0.107}} \\
		cDDPM     
		& 38.9 & 0.956 & \underline{\textbf{0.049}}
		& & 38.3 & 0.951 & 0.061
		& & 32.1 & 0.808 & \underline{\textbf{0.179}}
		& & 32.8 & 0.844 & 0.167 \\
		U\mbox{-}Net    
		& \underline{\textbf{41.6}} & \underline{\textbf{0.966}} & 0.110
		& & 39.2 & \underline{\textbf{0.963}} & 0.108
		& & \underline{\textbf{33.5}} &  \underline{\textbf{0.820}} & 0.285
		& & 32.9 & 0.849 & 0.209 \\
		EDSR      
		& 41.1 & 0.965 & 0.114
		& & \underline{\textbf{40.3}} & 0.954 & 0.104
		& & 33.2 & 0.818 & 0.283
		& & 33.1 & 0.849 & 0.202 \\
		GAN-CIRCLE 
		& 40.8 & 0.964 & 0.116
		& & 40.0 & 0.961 & 0.108
		& & 33.3 & 0.818 & 0.292
		& & 33.3 & \underline{\textbf{0.875}} & 0.177 \\
		ESRGAN    
		& 39.5 & 0.951 & 0.068
		& & 37.7 & 0.940 & 0.085
		& & 32.6 & 0.802 & 0.205
		& & 31.4 & 0.784 & 0.155 \\
		LR        
		& 33.5 & 0.915 & 0.228
		& & 34.3 & 0.936 & 0.160
		& & 29.7 & 0.761 & 0.389
		& & 31.1 & 0.860 & 0.219 \\
		\hline\hline
	\end{tabularx}
	\label{tab:metrics}
\end{table*}
\begin{figure*}[!t]
	\begin{center}
			\includegraphics[width=1\linewidth]{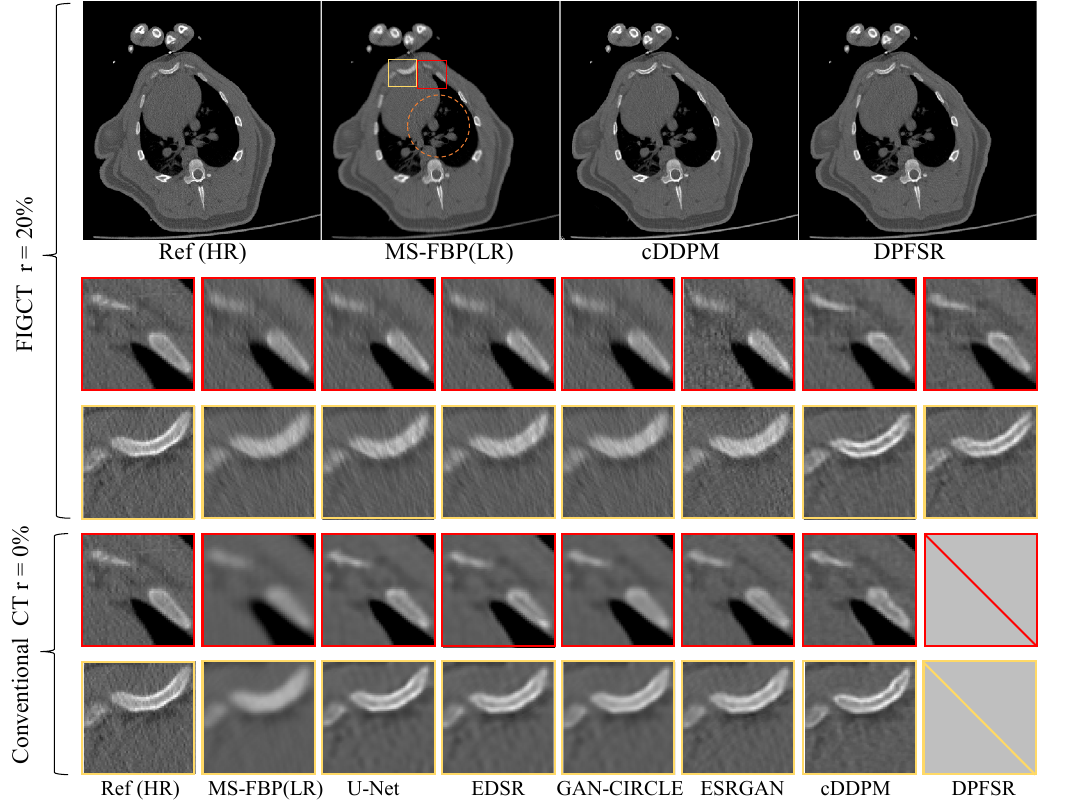}
	\end{center}
	\caption{ Visual comparison on a swine data for FIGCT $r = 20\%$ and conventional CT $r = 0\%$ cases across different methods. The display windows is [-500,1000] HU. Region within orange boxes are zoomed in.}
	\label{fig:result_exp} 
\end{figure*}

\subsection{Validation on Swine Data} 
A swine lung data was used to evaluate the performance of DPFSR. The phantom was scanned on a helical CT
platform. The scanning parameters are briefly summarized as follows: tube voltage 140kV, tube current 11mA, 720 projections over a range of 360 degrees, exposure time of 0.4 sec per rotation. We collected data from 10 rotations and averaged them to reduce image noise. The images were first reconstructed at high resolution and then forward-projected following the simulation procedure, yielding mixed-resolution projection of FIGCT and the corresponding reconstructed LR images.

\subsubsection{Comparison of DPFSR and Other Methods}
The quantitative results are summarized in Table~\ref{tab:metrics} for different methods. DPFSR achieves the best performance in both PSNR and LPIPS. This is consistent with the visual results in Fig.~\ref{fig:result_exp}, where DPFSR generates SR images with better structural preservation of bone and richer texture.

\subsubsection{Comparison of FIGCT and Conventional CT Geometry}
Compared with the FIGCT $r = 20\%$ case, the visual results in Fig.~\ref{fig:result_exp} show that SR images generated from the conventional CT $r = 0\%$ LR inputs generally deviate more substantially from the ground truth and exhibit noticeable oversmoothing. In particular, DDPM produces artificial structures that are not present in the target image. This suggests that the HR segment of projection in FIGCT substantially improves the generalization ability of the network.

\subsection{SR Performance on A-FIGCT}
To further investigate the effectiveness of DPFSR in other FIGCT architecture, we use A-FIGCT as example, in which detector works in different binning modes in different projection views.

We compared the SR performance of different methods on the A-FIGCT dataset with \(r = 10\%\) and on the conventional CT  \(r = 0\%\) dataset. DPFSR is modified to accommodate this architecture, while Eq.~(\ref{eq:M_image}) is formulated as follows without the mask operation.
\begin{equation}
	{\hat{x}_{0|t}} = \mathcal{R}({\hat{P}_{0|t}})
\end{equation}

\begin{figure*}[t]
	\begin{center}
			\includegraphics[width=1\linewidth]{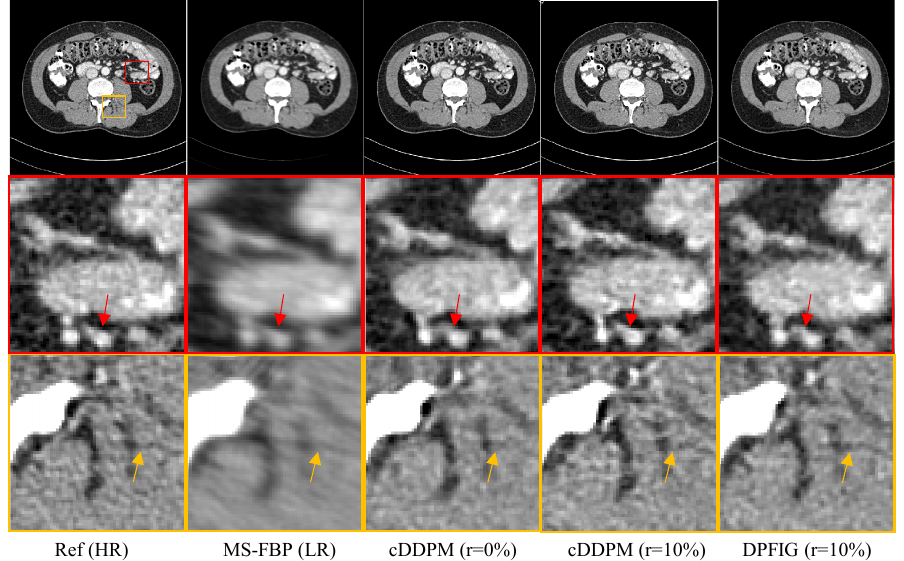}
	\end{center}
	\caption{Visual comparison of A-FIGCT $r = 10\%$ case and conventional CT $r = 0\%$ case . The display windows is [-160,240] HU. Region within yellow boxes are zoomed in. The regions indicated by the red and yellow arrows show that DPFSR ($r=20\%$) yields better results than cDDPM ($r=20\%$), cDDPM $r=0\%$.}
	\label{fig:result_LA} 
\end{figure*}

Results super-resolved by DPFSR are presented in Fig.~\ref{fig:result_LA}. From visual inspection, the SR results with A-FIGCT $r = 10\%$ recover more image details than those with conventional CT $r = 0 \%$. Moreover, at A-FIGCT $r = 10\%$ case, DPFSR produces results closer to the ground truth image than cDDPM. This is also consistent with the quantitative results, in which DPFSR achieves the best performance across all three metrics. The above results indicating that the proposed A-FIGCT architecture and the DPFSR method can effectively enhance super-resolution performance.
\begin{table}[t]
	\centering
	\caption{Quantitative comparison of SR performance on conventional CT $r = 0\%$  and A-FIGCT $r = 10\%$ cases.}
	\renewcommand{\arraystretch}{1.6}
	\footnotesize
	\begin{tabular*}{\linewidth}{@{\extracolsep{\fill}}lccc}
		\hline\hline
		& PSNR & SSIM & LPIPS \\
		\hline
		LR & 33.4 & 0.919 & 0.121 \\
		DDPM ($r=0\%$)  & 38.3 & \underline{\textbf{0.951}} & 0.055 \\
		DDPM ($r=10\%$) & 38.3  & 0.950  & 0.054 \\
		DPFSR ($r=10\%$) & \underline{\textbf{38.9}} & \underline{\textbf{0.951}} & \underline{\textbf{0.044}} \\
		\hline\hline
	\end{tabular*}
\end{table}

\section{Discussions and Conclusions}
In this study, we introduced the FIGCT architecture and the DPFSR method for global SR reconstruction. FIGCT simultaneously acquires LR projections covering the entire FOV and HR projections from the interior region, thereby providing local HR information while maintaining global anatomical coverage. This hybrid-resolution acquisition enables HR imaging in the interior region, which can be useful for CT applications with predefined or approximately localized regions of interest, such as cardiac CT, image-guided therapy, and dental CT. When combined with DPFSR, it further supports global SR reconstruction over the full FOV.

FIGCT employs a central HR detector surrounded by peripheral LR detector. This configuration substantially reduces hardware cost and data acquisition burden while preserving high image quality in regions of interest. Within the interior HR region, the reconstruction quality achieves PSNR values above 60 dB, comparable to fully HR imaging. Therefore, FIGCT provides a low-complexity and resource-efficient solution for fast HR ROI imaging.

To obtain global HR images, we further proposed DPFSR for FIGCT, in which HR information from the HR projection segment is incorporated into intermediate clean-image estimates in both the projection and image domains during the reverse diffusion process. Results on the test set show that DPFSR achieves the best overall performance, with a PSNR of 39.9, an SSIM of 0.959, and an LPIPS of 0.043. In the swine-data experiments, DPFSR also achieves the best PSNR and LPIPS, demonstrating improved structural fidelity, perceptual quality, and generalization across different objects.

\begin{figure*}[ht]
	\begin{center}
			\includegraphics[width=1\linewidth]{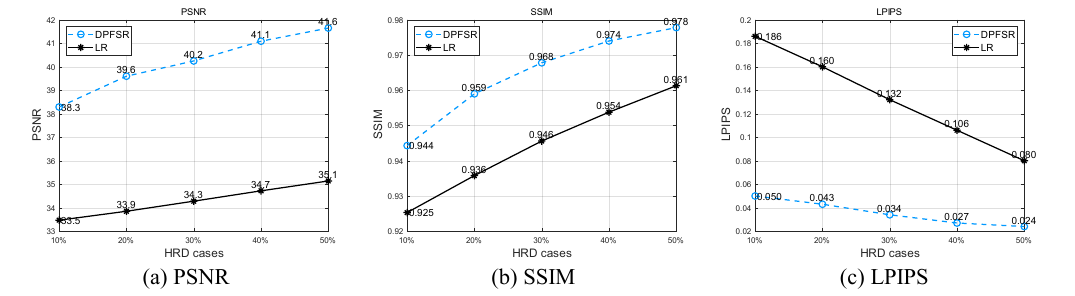}
	\end{center}
	\caption{Metrics variance of HDF $r$ from 10\% to 50\% cases initial reconstructed LR images and super resolving results.}
	\label{fig:ratio} 
\end{figure*}

To investigate the effect of the HDF ratio $r$ on the HR image generation capability of DPFSR, we trained and evaluated the model on five datasets with $r$ ranging from $10\%$ to $50\%$. As shown in Fig.~\ref{fig:ratio}, the PSNR, SSIM, and LPIPS values of the LR images vary approximately linearly with $r$, while those of the SR images also show a positive correlation with $r$. For both LR and SR images, PSNR and SSIM exhibit the most pronounced improvement when $r$ increases from $10\%$ to $20\%$, whereas LPIPS shows the largest decrease at $r = 30\%$. All metrics were calculated without applying Eq.~(\ref{eq:mask}).
Our results further highlight the importance of the FIGCT configuration for CT super-resolution. In the swine-data experiments, the model trained with FIGCT at $r = 20\%$ generates images with richer details and closer resemblance to the target images, owing to the presence of HR projection components. In contrast, the model trained on conventional CT data with $r = 0\%$ tends to produce overly smooth images with structural inconsistencies. Simulation results on FIGCT and A-FIGCT further demonstrate that DPFSR effectively integrates HR projection segments into the reconstruction process, achieving superior super-resolution performance compared with conventional CT-based SR methods.

FIGCT holds great promise for practical deployment across various CT system designs. In multi-row detector CT systems, detector elements with different physical sizes can be incorporated within a single detector array, as illustrated in Fig.~\ref{fig:FIGCT_example}. In flat-panel-detector-based cone-beam CT and photon-counting CT, FIGCT can be implemented by applying different binning modes to different detector regions, while A-FIGCT can be realized by using different binning modes across projection views, thereby reducing data transfer and storage requirements. In addition, heterogeneous detector types can also be integrated within the FIGCT framework. For example, Co-TeGRT~\cite{Xia20} combines kV-CT and PET detectors, enabling a more compact system design. Combining photon-counting detectors with energy-integrating detectors can improve system resolution while reducing overall system cost~\cite{PCD-EID}. Moreover, D-FIGCT can be implemented using either multi-row detectors or flat-panel detectors.


The proposed FIGCT architecture and  DPFSR method also has several limitations. First, DPFSR is currently designed for 2D slice-based reconstruction, and its extension to fully 3D reconstruction remains an important direction for future work. Second, the current FIGCT design provides HR imaging only for a predefined or approximately localized region, and therefore its applicability may be limited to task-specific CT scenarios, such as cardiac CT, image-guided radiation therapy, and dental CT, where the target region can be approximately determined before acquisition. Therefore, further investigation is needed to improve the robustness and reliability of DPFSR in practical imaging scenarios.

\clearpage

 
\section*{References}
\addcontentsline{toc}{section}{\numberline{}References}
\vspace*{-20mm}





\bibliography{./report.bib}      

\begin{thebibliography}{10}

\bibitem{Pontone14Radio}
G.~Pontone, E.~Bertella, S.~Mushtaq, M.~Loguercio, S.~Cortinovis, A.~Baggiano,
  E.~Conte, A.~Annoni, A.~Formenti, V.~Beltrama, A.~I. Guaricci, and
  D.~Andreini,
\newblock Coronary Artery Disease: Diagnostic Accuracy of CT Coronary
  Angiography—A Comparison of High and Standard Spatial Resolution Scanning,
\newblock Radiology {\bf 271}, 688--694 (2014),
\newblock PMID: 24520943.

\bibitem{yang2024empirical}
Y.~Yang, S.~Wang, D.~Pal, Z.~Yin, N.~J. Pelc, and A.~S. Wang,
\newblock Empirical optimization of energy bin weights for compressing
  measurements with realistic photon counting x-ray detectors,
\newblock Medical physics {\bf 51}, 224--238 (2024).

\bibitem{mp_resolution_impact}
A.~M. Hernandez, P.~Wu, M.~Mahesh, J.~H. Siewerdsen, and J.~M. Boone,
\newblock Location and direction dependence in the 3D MTF for a high-resolution
  CT system,
\newblock Medical Physics {\bf 48}, 2760--2771 (2021).

\bibitem{xiaochuan_pan_spatial-resolution_2005}
{Xiaochuan Pan}, {Lifeng Yu}, and {Chien-Min Kao},
\newblock Spatial-resolution enhancement in computed tomography,
\newblock IEEE Transactions on Medical Imaging {\bf 24}, 246--253 (2005),
\newblock Publisher: Institute of Electrical and Electronics Engineers (IEEE).

\bibitem{Zhao_2023}
Q.~Zhao, J.~Li, Y.~Li, and S.~Luo,
\newblock Super-resolution model-based iterative reconstruction for
  lens-coupled micro-CT imaging,
\newblock Measurement Science and Technology {\bf 34}, 085401 (2023).

\bibitem{haneda2022high}
E.~Haneda, B.~De~Man, B.~Claus, and L.~Fu,
\newblock High-resolution CT reconstruction from Zoom-In Partial Scans (ZIPS)
  with simultaneous estimation of inter-scan ROI motion,
\newblock in {\em Medical Imaging 2022: Physics of Medical Imaging}, volume
  12031, pages 646--653, SPIE, 2022.

\bibitem{prokop_general_2003}
M.~Prokop,
\newblock General principles of {MDCT},
\newblock European Journal of Radiology {\bf 45}, S4--S10 (2003).

\bibitem{stevenhigh}
I.~Steven~Tilley, W.~Zbijewski, and J.~W. Stayman,
\newblock High-Fidelity Modeling of Shift-Variant Focal-Spot Blur for
  High-Resolution CT,
\newblock Int’l Mtg. Fully 3D Image Recon. in Radiology and Nuc. Med .

\bibitem{tilley2015model}
S.~Tilley, J.~H. Siewerdsen, and J.~W. Stayman,
\newblock Model-based iterative reconstruction for flat-panel cone-beam CT with
  focal spot blur, detector blur, and correlated noise,
\newblock Physics in Medicine \& Biology {\bf 61}, 296 (2015).

\bibitem{zhong2021preliminary}
X.~Zhong, A.~Cai, N.~Liang, X.~Yu, L.~Li, and B.~Yan,
\newblock Preliminary studies on Dual-energy CT image super-resolution based on
  dual-dictionary learning,
\newblock in {\em 2021 IEEE International Conference on Medical Imaging Physics
  and Engineering (ICMIPE)}, pages 1--4, IEEE, 2021.

\bibitem{LI2016196}
J.~Li, J.~Wu, H.~Deng, and J.~Liu,
\newblock A self-learning image super-resolution method via sparse
  representation and non-local similarity,
\newblock Neurocomputing {\bf 184}, 196--206 (2016),
\newblock RoLoD: Robust Local Descriptors for Computer Vision 2014.

\bibitem{zhang2021self}
Z.~Zhang, S.~Yu, W.~Qin, X.~Liang, Y.~Xie, and G.~Cao,
\newblock Self-supervised CT super-resolution with hybrid model,
\newblock Computers in Biology and Medicine {\bf 138}, 104775 (2021).

\bibitem{chi2022ct}
J.~Chi, Z.~Sun, H.~Wang, P.~Lyu, X.~Yu, and C.~Wu,
\newblock CT image super-resolution reconstruction based on global hybrid
  attention,
\newblock Computers in Biology and Medicine {\bf 150}, 106112 (2022).

\bibitem{zhu2024super}
J.~Zhu, T.~Su, X.~Zhang, H.~Cui, Y.~Tan, H.~Zheng, D.~Liang, J.~Guo, and Y.~Ge,
\newblock Super-resolution dual-layer CBCT imaging with model-guided deep
  learning,
\newblock Physics in Medicine \& Biology {\bf 69}, 015016 (2024).

\bibitem{yu_computed_2017}
H.~Yu, D.~Liu, H.~Shi, H.~Yu, Z.~Wang, X.~Wang, B.~Cross, M.~Bramler, and T.~S.
  Huang,
\newblock Computed tomography super-resolution using convolutional neural
  networks,
\newblock in {\em 2017 {IEEE} {International} {Conference} on {Image}
  {Processing} ({ICIP})}, pages 3944--3948, Beijing, 2017, IEEE.

\bibitem{Park_SR}
J.~Park, D.~Hwang, K.~Y. Kim, S.~K. Kang, Y.~K. Kim, and J.~S. Lee,
\newblock Computed tomography super-resolution using deep convolutional neural
  network.,
\newblock Physics in Medicine \& Biology {\bf 63}, 145011 (2018).

\bibitem{DDPM}
J.~Ho, A.~Jain, and P.~Abbeel,
\newblock Denoising Diffusion Probabilistic Models,
\newblock in {\em Advances in Neural Information Processing Systems}, edited by
  H.~Larochelle, M.~Ranzato, R.~Hadsell, M.~Balcan, and H.~Lin, volume~33,
  pages 6840--6851, Curran Associates, Inc., 2020.

\bibitem{SR3}
C.~Saharia, J.~Ho, W.~Chan, T.~Salimans, D.~J. Fleet, and M.~Norouzi,
\newblock Image Super-Resolution via Iterative Refinement,
\newblock IEEE Transactions on Pattern Analysis and Machine Intelligence {\bf
  45}, 4713--4726 (2023).

\bibitem{SRDiff}
H.~Li, Y.~Yang, M.~Chang, S.~Chen, H.~Feng, Z.~Xu, Q.~Li, and Y.~Chen,
\newblock SRDiff: Single image super-resolution with diffusion probabilistic
  models,
\newblock Neurocomputing {\bf 479}, 47--59 (2022).

\bibitem{dolce}
J.~Liu, R.~Anirudh, J.~J. Thiagarajan, S.~He, K.~A. Mohan, U.~S. Kamilov, and
  H.~Kim,
\newblock {DOLCE}: {A} {Model}-{Based} {Probabilistic} {Diffusion} {Framework}
  for {Limited}-{Angle} {CT} {Reconstruction},
\newblock in {\em 2023 {IEEE}/{CVF} {International} {Conference} on {Computer}
  {Vision} ({ICCV})}, pages 10464--10474, Paris, France, 2023, IEEE.

\bibitem{zhang21}
X.~Zhang, V.~A. Kelkar, J.~Granstedt, H.~Li, and M.~A. Anastasio,
\newblock Impact of deep learning-based image super-resolution on binary signal
  detection,
\newblock Journal of Medical Imaging {\bf 8}, 065501--065501 (2021).

\bibitem{tivnan_hallucination_2024}
M.~Tivnan, S.~Yoon, Z.~Chen, X.~Li, D.~Wu, and Q.~Li,
\newblock Hallucination {Index}: {An} {Image} {Quality} {Metric} for
  {Generative} {Reconstruction} {Models},
\newblock in {\em Medical {Image} {Computing} and {Computer} {Assisted}
  {Intervention} – {MICCAI} 2024}, edited by M.~G. Linguraru, Q.~Dou,
  A.~Feragen, S.~Giannarou, B.~Glocker, K.~Lekadir, and J.~A. Schnabel, pages
  449--458, Cham, 2024, Springer Nature Switzerland.

\bibitem{CircleGAN}
C.~You, G.~Li, Y.~Zhang, X.~Zhang, H.~Shan, M.~Li, S.~Ju, Z.~Zhao, Z.~Zhang,
  W.~Cong, M.~W. Vannier, P.~K. Saha, E.~A. Hoffman, and G.~Wang,
\newblock CT Super-Resolution GAN Constrained by the Identical, Residual, and
  Cycle Learning Ensemble (GAN-CIRCLE),
\newblock IEEE Transactions on Medical Imaging {\bf 39}, 188--203 (2020).

\bibitem{luo_interior_2018}
S.~Luo, T.~Shen, Y.~Sun, J.~Li, G.~Li, and X.~Tang,
\newblock Interior tomography in microscopic {CT} with image reconstruction
  constrained by full field of view scan at low spatial resolution,
\newblock Physics in Medicine \& Biology {\bf 63}, 075006 (2018),
\newblock Publisher: IOP Publishing.

\bibitem{Xia20}
H.~Gao and Y.~Xia,
\newblock A novel coplanar multi-modality tomographic imaging for image
  guidance in radiotherapy using hybrid radiation detector,
\newblock in {\em MEDICAL PHYSICS}, volume~47, pages E570--E570, WILEY 111
  RIVER ST, HOBOKEN 07030-5774, NJ USA, 2020.

\bibitem{Zhang_2025}
Z.~Zhang, B.~Chen, D.~Xia, E.~Y. Sidky, and X.~Pan,
\newblock Accurate image reconstruction within and beyond the field-of-view of
  CT system from data with truncation,
\newblock Physics in Medicine \& Biology {\bf 70}, 035005 (2025).

\bibitem{Repainted}
A.~Lugmayr, M.~Danelljan, A.~Romero, F.~Yu, R.~Timofte, and L.~Van~Gool,
\newblock {RePaint}: {Inpainting} using {Denoising} {Diffusion} {Probabilistic}
  {Models},
\newblock in {\em 2022 {IEEE}/{CVF} {Conference} on {Computer} {Vision} and
  {Pattern} {Recognition} ({CVPR})}, pages 11451--11461, New Orleans, LA, USA,
  2022, IEEE.

\bibitem{xia2021magic}
W.~Xia, Z.~Lu, Y.~Huang, Z.~Shi, Y.~Liu, H.~Chen, Y.~Chen, J.~Zhou, and
  Y.~Zhang,
\newblock MAGIC: Manifold and Graph Integrative Convolutional Network for
  Low-Dose CT Reconstruction,
\newblock IEEE Transactions on Medical Imaging {\bf 40}, 3459--3472 (2021).

\bibitem{LPIPS}
R.~Zhang, P.~Isola, A.~A. Efros, E.~Shechtman, and O.~Wang,
\newblock The {Unreasonable} {Effectiveness} of {Deep} {Features} as a
  {Perceptual} {Metric},
\newblock in {\em 2018 {IEEE}/{CVF} {Conference} on {Computer} {Vision} and
  {Pattern} {Recognition}}, pages 586--595, Salt Lake City, UT, 2018, IEEE.

\bibitem{LDCT}
C.~McCollough,
\newblock 2016 Low-Dose CT Grand Challenge, 2022.

\bibitem{U-net}
O.~Ronneberger, P.~Fischer, and T.~Brox,
\newblock U-net: Convolutional networks for biomedical image segmentation,
\newblock in {\em International Conference on Medical image computing and
  computer-assisted intervention}, pages 234--241, Springer, 2015.

\bibitem{EDSR}
B.~Lim, S.~Son, H.~Kim, S.~Nah, and K.~Mu~Lee,
\newblock Enhanced deep residual networks for single image super-resolution,
\newblock in {\em Proceedings of the IEEE conference on computer vision and
  pattern recognition workshops}, pages 136--144, 2017.

\bibitem{esrgan}
X.~Wang, K.~Yu, S.~Wu, J.~Gu, Y.~Liu, C.~Dong, Y.~Qiao, and C.~Change~Loy,
\newblock Esrgan: Enhanced super-resolution generative adversarial networks,
\newblock in {\em Proceedings of the European conference on computer vision
  (ECCV) workshops}, pages 0--0, 2018.

\bibitem{PCD-EID}
K.~Taguchi, G.~S.~K. Fung, Q.~Tang, and J.~Cammin,
\newblock {Hybrid EID algorithm for PCD/EID-CT systems},
\newblock in {\em Medical Imaging 2013: Physics of Medical Imaging}, edited by
  R.~M. Nishikawa, B.~R. Whiting, and C.~Hoeschen, volume 8668, page 86682M,
  International Society for Optics and Photonics, SPIE, 2013.

\end{thebibliography}



\bibliographystyle{./medphy.bst}    


\end{document}